\documentclass[12pt]{iopart}
\usepackage{epsfig}
\usepackage{iopams}
\newcommand{\lov}{\frac{l}{2}}

\newcommand{\sumksig}{\sum\limits_{k,\; \sigma}}

\newcommand{\ccd}{c_{-k' \downarrow}c_{k' \uparrow}}
\newcommand{\ccdag}{c^{\dag}_{k\uparrow}c^{\dag}_{-k\downarrow}}

\newcommand{\avs}[1]{ {\langle #1 \rangle} }

\begin{document}

\title[Charge qubits from a many-body perspective]{Superconducting charge qubits from a microscopic
many-body perspective}%
\author{D.A. Rodrigues\dag}
\address{School of Physics and Astronomy, University of
Nottingham, Nottingham, NG7 2RD, U.K.}%
 \author{T.P. Spiller}
\address{Hewlett Packard Laboratories, Filton Road, Bristol, BS34
8QZ, U.K.}
\author{ J.F. Annett and B.L. Gy\"orffy}
\address{Department of Physics, Bristol University, Bristol, BS8 1TL, U.K.}%

\ead{\mailto{\dag denzil.rodrigues@nottingham.ac.uk}}

\begin{abstract}
The quantised Josephson junction equation that underpins the
behaviour of charge qubits and other tunnel devices is usually
derived through cannonical quantisation of the classical macroscopic
Josephson relations. However, this approach may neglect effects due
to the fact that the charge qubit consists of a superconducting
island of finite size connected to a large superconductor.
 We show that the well known
quantised Josephson equation can be derived directly and simply from
a microscopic many-body Hamiltonian. By choosing the appropriate
strong coupling limit we produce a highly simplified Hamiltonian
that nevertheless allows us to go beyond the mean field limit and
predict further finite-size terms in addition to the basic equation.
\end{abstract}
\submitto{JPhysC}

\maketitle

The Josephson effect \cite{josephson} is still one of the phenomena
that make superconductors such a fascinating area of study. For
instance, after some 40 years of intensive basic and applied
research there are still new features of the coherent tunneling of
Cooper Pairs coming to light in connection with junctions involving
small superconducting grains. The new complications arise from the
fact that when one side of a Josephson junction is sufficiently
small for the charging energy to be relevant, quantum interplay
between charging and tunneling begins to appear \cite{tinkham}. The
standard approach to describing these junctions is to take the
classical equations of motion of the superconducting phase
difference $\phi_D$ across the junctions and apply cannonical
quantisation rules to $\phi_D$ as a `position' variable. Of course,
the classical Josephson equations for such a phase difference are
first
 derived from microscopic theory \cite{josephson}, and so this
standard approach represents a `re-quantisation' of `classical'
equations that were in turn derived from quantum mechanical
microscopic theory using a mean field theory that does not take into
account the charging energy of the island. Consequently, in such
approaches the description of quantum fluctuations is at best
semi-phenomenological. In what follows we examine the limitations of
the above procedure on the basis of a simple model which permits an
exact treatment of a superconducting island coupled weakly to a bulk
superconductor.

To motivate our interest in the problem we note that much current
experimental and theoretical attention is focused on nanoscale
superconducting grains coupled to large superconductors as such
`Cooper Pair Boxes' are becoming realistic candidates for being
useful qubits in Quantum Information devices
\cite{shnirman97,shnirmannature,shnirmanreview,younori}. Starting
with the work of Nakamura et al. \cite{nakamura}, over the last few
years there have been a number of impressive experiments
\cite{quantronium,nakamura2, nakamura3, delsing,jin, savelev}
demonstrating appropriate charge qubit behaviour and macroscopic
tunnelling. As experiments continue to improve, it is now pertinent
to re-examine the standard approach to describing quantum
fluctuations in superconducting charge qubits and related systems.
Clearly, in the course of such investigation one would expect to
reproduce the basic quantum phenomenology from a fully microscopic
approach within a well controlled approximation, as there is already
good experimental support for this. Nevertheless, a generalised
theory will also yield new additional terms due to the finite size
of the superconducting islands and  for future experiments they
could have significant consequences.  To shed light on these, we
examined a simple microscopic model of a Cooper Pair box, showing
how the familiar phenomenology emerges, along with new finite size
effects.

\section{Quantising the Josephson Relations}

In the interest of clarity, we start our discussion by recalling,
briefly, the usual phenomenological approach to the problem at hand.
The standard way to obtain the Hamiltonian describing a small
superconductor connected through Josephson tunnel junctions to a
bulk superconductor is by starting with the Josephson equations for
$\phi_D$ and $V$, the difference in phases of the two
superconducting regions and the voltage across the junction,
\begin{eqnarray}\label{eq:qjj1}
I &=& I_C \sin{\phi_D} \\
\frac{d \phi_D}{dt} &=& \frac{2eV}{\hbar} \label{eq:qjj2}
\end{eqnarray}
where $I_C$ is the critical current of the junction. Although
derived from a quantum mechanical microscopic treatment, as
evidenced by the appearance of $\hbar$, these equations can be
regarded as classical equations of motion. We follow the standard
procedure for canonical quantisation and first find the Lagrangian
that leads to these equations. Namely, we take,
\begin{eqnarray}\label{eq:scqu-lagrange}
\mathcal{L} = \frac{1}{2}  \frac{\hbar^2 C}{4 e^2} \left( \frac{d
\phi_D}{d t} \right)^2 + \frac{\hbar I_C}{2 e} \cos{\phi_D}
\end{eqnarray}
where we have introduced the total island capacitance, $C$. If we
choose the phase $\phi_D$ to be the canonical position variable, we
can identify the canonical momentum $\pi=\partial
\mathcal{L}/\partial \dot{\phi_D}$,
\begin{eqnarray}
\pi &=& \frac{\hbar^2 C}{4 e^2} \; \dot{\phi_D}= \frac{\hbar \;
CV}{2e}  = \hbar (N-n_g),
\end{eqnarray}
and it is seen that the phase of the condensate and the excess
number of Cooper pairs on the island, $(N-n_g)$, are conjugate
variables \cite{shnirman97,shnirmannature}. The term $n_g$
represents an applied gate voltage (in dimensionless units) and so
the canonical momentum $\pi=(N-n_g)$ is effectively the charge on
the device, viewed as a capacitor, in units of $2 e$. Note that if
the total charge on the system is zero, this can be rewritten in
terms of the charge difference. To quantise the system, we introduce
the commutation relation between conjugate variables $[\phi_D,
\pi]=i\hbar$ and note that this can be satisfied by writing
$\pi=-i\hbar \;\partial/\partial\phi_D-n_g$ (keeping the gate
voltage explicit). Then the Hamiltonian,
$\widehat{H}=\pi\dot{\phi_D}-\mathcal{L}$, is given by,
\begin{eqnarray}\label{eq:scqu-qmham}
\widehat{H} &=& E_C  \left(
i\frac{\partial}{\partial\phi_D}-n_g\right)^2 - E_J \cos{{\phi_D}}.
\end{eqnarray}
where the charging energy is given by $E_C=2e^2/C$ and $E_J$ is
defined as $E_J=\hbar I_C/2e$. The Schr\"odinger's equation for the
amplitude $\psi(\phi_D)$ is then,
\begin{eqnarray}\label{eq:scqu-qmham}
\widehat{H} \psi(\phi_D) &=& i \hbar\frac{d}{dt} \psi(\phi_D).
\end{eqnarray}
Evidently, the probability that the phase difference takes on a
certain value is given by $|\psi(\phi_D)|^2$.

This is the desired standard quantum description of the Josephson
Junction. In the remainder of this paper, we show how the above
quantised Josephson junction equation can be rederived directly from
the microscopic theory in a way that includes finite size effects.

\section{Finite Superconductors As Spins}

We wish to produce a description of a finite superconductor that is
simple enough to solve exactly but retains properties due to its
finite size. Specifically, we wish to be able to capture effects
that go beyond the mean field approximation. Our starting point is
the well-known BCS Hamiltonian \cite{BCS}:

\begin{equation} \label{eq:BCSfull}
\hat{H} = \sumksig{\epsilon_k c^{\dag}_{k,\; \sigma} c_{k,\; \sigma}
} - \sum\limits_{  k,\; k' } V_{k,\; k'} \; \ccdag \ccd,
\end{equation}
where ${c_{k,\;\sigma }^{\dag }}$ and ${c_{k,\;\sigma }}$ create and
annihilate electrons, respectively, with spin $\sigma $ in the state
$k$ with energy ${\epsilon _{k}}$ and the matrix element $
V_{k,\;k^{\prime }}$ describes an attractive two body interaction.
As we are discussing a finite superconducting island, the label
$k,\uparrow$ does not refer to a free electron wavevector but to a
generic single-electron eigenstate, with $-k,\downarrow$
representing the corresponding time-reversed state.

\par A common approximation to this equation is made by
assuming the pairing potential $V_{k,\; k'}$ is equal for all $k,\;
k'$ in a region around the Fermi energy determined by the cutoff
energy $\hbar \omega_c$ and zero outside this region. That is,
$V_{k,\; k'}=V$ for $|\epsilon_k-\epsilon_F|<\hbar \omega_c$ and
$V_{k,\; k'}=0$ otherwise. This greatly simplifies matters whilst
retaining the essential physics. We now adopt a similar philosophy
in making a further approximation, and take all the single electron
energy levels $\epsilon_k$ within the cutoff region around the Fermi
energy to be equal to the Fermi energy $\epsilon_F$.

The interaction term, $V_{k,k'}$ acts only within the cutoff region
around the Fermi energy. Outside this region the Hamiltonian is
diagonal and trivially solved. Writing the single electron energy as
$\epsilon_k= \epsilon_F+(\epsilon_k-\epsilon_F)$, we note that
within the cutoff region $|\epsilon_k-\epsilon_F|<\hbar \omega_c$,
and thus if $V\gg\hbar \omega_c$, then $|\epsilon_k-\epsilon_F| \ll
V$ and we can discard the variation of $\epsilon_k$. Thus in this
strong coupling approximation, our Hamiltonian becomes,
\begin{equation} \label{eq:BCSscl}
    \widehat{H} =  \epsilon_F\sum\limits_{k}' c^{\dag}_{k,\;\sigma}c_{k,\;\sigma}
     - V \sum\limits_{ k, \; k'}'
\ccdag \ccd
\end{equation}
where the dashes on the sums indicate that they are only taken
over states within the cutoff region.

Although this caricature of a realistic Hamiltonian represents an
uncontrolled approximation, we will show that it allows us to derive
a Josephson junction equation that goes beyond mean field, and that
the results it produces agree in the strong coupling limit with
known results in two important cases. Namely, the mean-field
solution of this Hamiltonian agrees with the BCS solution, and the
exact solution agrees with the Richardson solution\cite{richardson}.

It should be noted that although many superconductors can be
described as having strong coupling, the BCS Hamiltonian is not
necessarily appropriate for their description. Equation
\ref{eq:BCSscl} is hence not intended as a description of this
particular class of superconductors, but rather as a generic model
that, although simplified, allows an exact solution and a
description of the physics we are trying to capture.

As a consequence of the above simplifications equation
\ref{eq:BCSscl} can now be written in terms of the three operators
\cite{leeandscully},
\begin{eqnarray} \label{eq:defn_s}
 \widehat{S}^{Z} &=&\frac{1}{2}\sum\limits_{k}' \left({c^{\dag}_{k\uparrow}c_{k\uparrow}+
    c^{\dag}_{-k\downarrow}c_{-k\downarrow}-1} \right) \nonumber \\
\widehat{S}^+ &=& \sum\limits_k'
c^{\dag}_{k\uparrow}c^{\dag}_{-k\downarrow} \nonumber\\
\widehat{S}^{-} &=&\sum\limits_{k}^{\prime }c_{-k\downarrow
}^{{}}c_{k\uparrow }^{{}},
\end{eqnarray}
Note that these operators obey the commutation relations, and
therefore the algebra, of quantum spin operators of size $l/2$,
where $l$ is the number of levels in the cutoff region. Thus the
main result of this section is the effective Hamiltonian,
\begin{equation} \label{eq:spinsc_H}
\widehat{H}_{sp} = 2(\epsilon_F-\mu)
\left(\widehat{S}^Z+\frac{l}{2}\right) -
V{\widehat{S}^+}{\widehat{S}^-},
\end{equation}
where we have introduced  a chemical potential $\mu $ to describe
coupling to a reservoir.

\section{Exact solution}

The eigenstates of equation \ref{eq:spinsc_H} are the eigenstates of
the spin operator $\widehat{S}^Z$, $|\frac{l}{2},m_N\rangle$, where
the component of the spin along the $Z$ axis is given by $m_N=N-l/2$
and $N$ denotes the number of Cooper pairs on the island. The
eigenenergies corresponding to these eigenstates are,
\begin{eqnarray}
E_N&=& 2(\epsilon_F-\mu)N-VN(l-N+1) \label{eq:energyN}.
\end{eqnarray}
A chemical potential allows us to specify the average number of
Cooper Pairs on the island in equilibrium. In the case of the exact
solution, where $N$ is a good quantum number, this means choosing a
state with a particular value of $N$ to be the ground state. We
choose $\mu$ so that the ground state is the state with a chosen
value of $N$, which we label $\bar{N}$. The eigenenergies $E_N$
therefore become,
\begin{eqnarray}
E_{N}&=& -VN(2\bar{N}-N)\label{eq:energywmu},
\end{eqnarray}
and the ground state is $|\frac{l}{2},m_{\bar{N}}\rangle$, with
energy,
\begin{eqnarray}
E_{gs}&=& -V\bar{N}^2\label{eq:energygs}.
\end{eqnarray}
We can also easily see that although the pairing parameter $\langle
\widehat{S}^+\rangle=0$ in all eigenstates, i.e. there is no
symmetry breaking, we still have fluctuations as expected for a
finite superconductor  which are given by $\langle
\widehat{S}^+\widehat{S}^-\rangle=N(l-N+1)$ for a general eigenstate
$N$ and
\begin{eqnarray} \label{eq:flucgs}
\langle\textrm{$\frac{l}{2}$},m_{\bar{N}}|
\widehat{S}^+\widehat{S}^-|\textrm{$\frac{l}{2}$},m_{\bar{N}}\rangle=\bar{N}(l-\bar{N}+1)
\end{eqnarray}
for the ground state. We also see that the operator $\widehat{S}^+$
which couples eigenstates, corresponds (when appropriately
normalised) to the quasiparticle creation operator for the system.

\section{Comparison to Standard Results}

To generate confidence in this simple model, we compare its
solutions to the solutions of the full Hamiltonian (equation
\ref{eq:BCSfull}) in two ways. First, we find the mean field
solution and compare it to the BCS results. Second, we compare the
exact solution to the Richardson solution in the appropriate limit.

\subsection{The Mean Field Solution}

The mean field approximation arises from the assumption that the
operators $\widehat{S}^\pm$ remain close to their expectation
values, $\avs{\widehat{S}^\pm}$. We write  $\widehat{S}^\pm =
\langle \widehat{S}^\pm \rangle + (\widehat{S}^\pm - \langle
\widehat{S}^\pm \rangle)$, and discard terms to second order or
higher in $(\widehat{S}^\pm - \langle \widehat{S}^\pm \rangle)$.
Writing $V\langle \widehat{S}^- \rangle= \Delta$, we find the mean
field Hamiltonian for our model,
\begin{eqnarray} \label{eq:hmf1}
\widehat{H}_{MF} &=& \ 2(\epsilon_F-\mu)
\left(\widehat{S}^Z+\frac{l}{2}\right) - \Delta \widehat{S}^+ -
\Delta^* \widehat{S}^- +\frac{|\Delta|^2}{V}
\end{eqnarray}
Apart from the constant term, this is a linear combination of the
spin operators $\widehat{S}^Z$, $\widehat{S}^Y$ and $\widehat{S}^X$
and is therefore proportional to the projection of a spin operator
on an unknown direction specified by the unit vector $\hat{n}$. Thus
denoting $\widehat{S}.\hat{n}$ by $\widehat{S}^Z_{\hat{n}}$, we may
write,
\begin{eqnarray} \label{eq:hmf2}
\widehat{H}_{MF} &=&\gamma \widehat{S}^Z_{\hat{n}}
+\frac{|\Delta|^2}{V}+2(\epsilon_F-\mu)\frac{l}{2},
\end{eqnarray}
and therefore the problem of diagonalising equation \ref{eq:hmf1} is
equivalent to the problem of rotating the axis of quantisation for
our effective spin operators. Requiring that the commutation
relations $[\widehat{S}^Z_{\hat{n}} , \widehat{S}^+_{\hat{n}} ]  =
\widehat{S}^+_{\hat{n}} \nonumber $ hold for spin operators in the
frame of reference where the axis of quantisation is along
$\widehat{n}$  determines both an expression for
$\widehat{S}^+_{\hat{n}}$ and the value of $\gamma$,
\begin{eqnarray} \label{eq:sn+2}
\widehat{S}^+_{\hat{n}} &=& \frac{2\Delta}{\gamma} \left(\widehat{S}^Z -\frac{\Delta}{2\xi_F -\gamma} \widehat{S}^+ - \frac{\Delta^*}{2\xi_F +\gamma} \widehat{S}^-\right)  \nonumber \\
\gamma &=& 2\sqrt{(\xi_F) ^2 + |\Delta|^2}
\end{eqnarray}
where $\xi_F=\epsilon_F-\mu$ and we note that $\gamma=2E_F$, the
energy of a Cooper Pair evaluated at the Fermi energy. The ground
state of our Hamiltonian can now be trivially found, as it
corresponds to the $m=-l/2$ eigenstate of the operator
$\widehat{S}^Z_{\hat{n}}$. Recalling that a maximal $m$ state of a
spin operator pointing in one direction is a spin coherent state
\cite{coherent} in any other we find,
\begin{eqnarray} \label{eq:spcoherent}
|\alpha\rangle &=&
\frac{1}{\sqrt{(1+|\alpha|^2)^l}}\sum\limits_{N=0}^l
 \frac{(\alpha^*\widehat{S}^+)^N|\textrm{$\frac{l}{2}$}, m_0\rangle}{N!} \nonumber \\
 &=& \frac{1}{\sqrt{(1+|\alpha|^2)^l}}\prod\limits_k ( 1 + \alpha^* \; \ccdag) |\textrm{$\frac{l}{2}$},m_0 \rangle
\end{eqnarray}
As one might expect, the second line is a way of writing the BCS
ground state wavefunction in the limit where all the levels have
equal probability of occupation, i.e. $u_k/v_k=\alpha^*$ for all
$k$. Making use of equation \ref{eq:hmf2} and the fact that
$\widehat{S}^-_{\hat{n}} |\alpha\rangle=0$ for the ground state, we
find that $\alpha=(\xi_F-E_F)/\Delta$. To complete the calculation,
we need to self-consistently determine the values $\Delta$ and
$\mu$, which is relatively simple in the spin model and gives,
\begin{eqnarray}\label{eq:spinNdel1}
|\Delta|^2 &=& V^2\bar{N}(l-\bar{N}) \\
\epsilon_{F} - \mu &=& V (l/2 -\bar{N})  \label{eq:spinNdel2}\\
H_{SMF}|\alpha \rangle&=&-V\bar{N}^2  |\alpha \rangle
\label{eq:MFenergy}
\end{eqnarray}
where $\bar{N}$ is the average occupation of the island. We find
equations \ref{eq:spcoherent}- \ref{eq:MFenergy} are exactly equal
to the expressions found if we were to solve the full BCS equation
and then take the weak coupling limit (this is shown in the appendix
of \cite{revival}). Comparing the mean field solution to the exact
solution, we see that, surprisingly, the exact (equation
\ref{eq:energygs}) and mean field (equation \ref{eq:MFenergy})
ground state energies are identical. However, we have a non-zero
pairing parameter $\langle S^- \rangle = \Delta/V$ and the
expectation value of the mean field coupling term,
\begin{eqnarray}
\Bigg\langle\alpha \Bigg| - \Delta \widehat{S}^+ - \Delta^*
\widehat{S}^- +\frac{|\Delta|^2}{V}\Bigg|\alpha\Bigg\rangle &=&
-V\bar{N}(l-\bar{N}),
\end{eqnarray}
neglects the quantum fluctuations present in the exact solution,
\begin{eqnarray}
\left\langle\frac{l}{2},m_{\bar{N}}\right|-V
\widehat{S}^+\widehat{S}^-\left|\frac{l}{2},m_{\bar{N}}\right\rangle&=&-V\bar{N}(l-\bar{N}+1)
\label{eq:excouple}
\end{eqnarray}
in much the same way a classical spin neglects the fluctuations
present in a quantum spin, i.e. the eigenvalues of $\widehat{S}^2$
are $S(S+1)$ and not the classical values $S^2$.

\subsection{The Richardson Solution}

Unbeknownst to the condensed matter community for many years, there
exists an exact solution to the BCS Hamiltonian (equation
\ref{eq:BCSfull}) for finite superconductors, first discovered in
1963 by Richardson\cite{richardson,delftsmallreview,ricrev} in the
context of nuclear physics. It has been shown that this solution
reproduces the BCS result in the bulk limit, but it is difficult to
work with for any island occupied by more than a few Cooper pairs.
The Richardson solution requires the introduction of operators that
diagonalise the full (i.e. not mean field) BCS Hamiltonian,
\begin{eqnarray}
\widehat{H} &=& \sum\limits_{\nu=1}^N E_{J\nu}
\widehat{B}^+_{J\nu}\widehat{B}^-_{J\nu}\label{eq:ric-diagH}\\
\widehat{B}_{J\nu}^+&=&\sum_k
\frac{\ccdag}{2\epsilon_k-E_{J\nu}},\label{eq:ric-defB}
\end{eqnarray}
where the sum in equation \ref{eq:ric-diagH} runs over $\nu$ up to
the total number of Cooper pairs on the island. The parameters
$E_{J\nu}$ are found by solving the equations,
\begin{eqnarray} \label{eq:ric-eqn}
1+\frac{2V}{ E_{J\eta}- E_{J\nu}} =V\sum\limits_k
\frac{1}{2\epsilon_k - E_{J\nu}} ,
\end{eqnarray}
for all $\nu$.

Whilst the usual BCS theory has an essential singularity at $V=0$,
the theory is well behaved near $1/V \sim 0$. Thus, following
Altshuler \emph{et al.},\cite{altshuler} we expand equation
\ref{eq:ric-eqn} in powers of $\hbar \omega_c/V$. Using $\epsilon_k
\sim \epsilon_F$ leads to,
\begin{eqnarray} \frac{1}{V}
+\sum\limits_{\nu=1}^N \frac{2}{ E_{J\eta}- E_{J\nu}}
=\frac{l}{(E_{J\eta} - 2\epsilon_F)} + \sum\limits_{k=1}^l
\frac{2(\epsilon_k-\epsilon_F)}{(E_{J\eta} - 2\epsilon_F)^2}.
\end{eqnarray}
Now discard the second term on the right as negligible, multiply by
$E_{J\eta} - 2\epsilon_F$, and sum over the $N$ parameters
$E_{J\eta}$ to obtain,
\begin{eqnarray} \label{eq:SSC-riceqn3}
\sum\limits_{\eta=1}^N \frac{E_{J\eta}-2\epsilon_F}{V} +
 N(N-1)+0 &=& Nl
\end{eqnarray}
where the double sums over $\eta, \nu$ have either vanished, or gone
to $N(N-1)$ due to symmetry. Finally, we recall that the energy of
the Richardson solution is given by a sum over $E_{J\eta}$, and
rewrite equation \ref{eq:SSC-riceqn3} to get the energy of an island
containing N Cooper pairs:
\begin{eqnarray} \label{eq:SSC-riceqn4}
E_N =  2\epsilon_F N- V N(l-N+1).
\end{eqnarray}
As heralded in the introduction this result matches the exact energy
of the spin Hamiltonian as given in equation \ref{eq:energyN} for
$\mu=0$.

Thus we have shown that although we have made a significant
approximation to the Hamiltonian, the results thereby derived are
consistent with results obtained by solving the full system in
either the mean field approximation or exactly, and \emph{then}
taking the appropriate limit.

\section{Phase Representation of the Spin Operators $\widehat{S}^+$ and $\widehat{S}^-$}

The preceding sections have established our model of a finite
superconducting system as a large spin, as given in equation
\ref{eq:spinsc_H}. We shall now go on to show how this model can be
used to derive a phase-representation description of the Josephson
effect in a system comprising a small island coupled to a larger
piece of bulk superconductor.

We wish to convert to a representation in terms of the continuous
phase variable $\phi$, i.e. convert from ket notation to
wavefunction $\psi(\phi)$ and differential operator (such as
$\frac{d}{d \phi}$) notation. Thus, a ket $|\psi_a\rangle$ will
become a wavefunction $\langle\phi|\psi_a\rangle$, and the
differential operator must be consistent with this. Defining the
state $|\phi\rangle=(2\pi)^{-1/2}\sum e^{i\phi
N}|\textrm{$\frac{l}{2}$},m_N\rangle$, we find that the wavefunction
corresponding to $|\textrm{$\frac{l}{2}$},m_N\rangle$ is
$(2\pi)^{-1/2}e^{-i\phi N}$. We can then examine how the operators
act on this wavefunction.
\begin{eqnarray}
\widehat{S}^Z\langle\phi|\textrm{$\frac{l}{2}$},m_N\rangle&=&\langle\phi|
(N-\textrm{$\frac{l}{2}$})|\textrm{$\frac{l}{2}$},m_N\rangle\nonumber\\
&=&(N-\textrm{$\frac{l}{2}$})\langle\phi|\textrm{$\frac{l}{2}$},m_N\rangle\nonumber\\
&=&(N-\textrm{$\frac{l}{2}$})\frac{e^{-i\phi N}}{\sqrt{2\pi}}\nonumber\\
\widehat{S}^Z\psi(\phi)&=&\left( i\frac{\partial}{\partial\phi} -
\frac{l}{2} \right)\psi(\phi)
\end{eqnarray}
Similarly, we find for the raising operator,
\begin{eqnarray}
\widehat{S}^+\langle\phi|\textrm{$\frac{l}{2}$},m_N\rangle&=&\langle\phi|
\sqrt{(N+1)(l-N)}|\textrm{$\frac{l}{2}$},m_N\rangle\nonumber\\
\widehat{S}^+\psi(\phi)&=&e^{-i\phi}\sqrt{\left(
i\frac{\partial}{\partial\phi}+1\right)\left(l-i\frac{\partial}{\partial\phi}
\right) }\;\psi(\phi)
\end{eqnarray}
with the lowering operator given by,
\begin{eqnarray}
\widehat{S}^-\langle\phi|\textrm{$\frac{l}{2}$},m_N\rangle&=&\langle\phi|
\sqrt{N)(l-N+1)}|\textrm{$\frac{l}{2}$},m_N\rangle\nonumber\\
\widehat{S}^-\psi(\phi)&=&e^{i\phi}\sqrt{
i\frac{\partial}{\partial\phi}\left(l-i\frac{\partial}{\partial\phi}
+1\right) }\;\psi(\phi)
\end{eqnarray}
Collecting the differential forms for the operators and rewriting
$S^+$ and $S^-$ into a more convenient form leaves us with,
\begin{eqnarray} \label{eq:phaseOP}
\widehat{S}^Z &=& \left( i\frac{\partial}{\partial\phi} - \frac{l}{2} \right) \nonumber\\
\widehat{S}^\pm &=&  \sqrt{\left(
\lov\pm\left(i\frac{\partial}{\partial\phi}-\lov \right)
\right)}e^{\mp i \phi} \sqrt{\left(
\lov\mp\left(i\frac{\partial}{\partial\phi}-\lov\right) \right)} \nonumber \\
\end{eqnarray}
The form of the raising and lowering operators can also be derived
by requiring that the commutation relations for quantum spins are
enforced. We see that it is the
$\left(i\frac{\partial}{\partial\phi}-\lov\right)$ terms in the
$S^+,S^-$ operators that take into account the finite size effects
and ensure that $[S^+,S^-]\neq0$.

Writing the operators in this form allows us to take the large size
($l\to \infty$) limit. In taking this limit we assume that
$S^Z\ll\lov$. In the superconducting language, this corresponds to
only states close to half filling being occupied. Specifically, we
assume,
\begin{eqnarray}
| \langle\textrm{$\frac{l}{2}$},m_N|\psi \rangle| \sim 0\; {for} \;
|N-\textrm{$\frac{l}{2}$}|\gtrsim
\left(\frac{l}{2}\right)^{\frac{1}{2}},
\end{eqnarray}
a condition which is fulfilled for coherent states with $\bar{N}$
set close to $\lov$. When this is true, we can expand the square
root in $(i\frac{\partial}{\partial\phi}-\lov)/l$. We see that the
leading order terms give $S^{\pm}=\lov e^{\mp i \phi}$, and we
regain the semiclassical large-size limit for which $[S^+,S^-]=0$ as
discussed by Lee and Scully\cite{leeandscully}.

\section{Quantised Josephson Junction Equation}

We can now use the forms of the operators given in equations
\ref{eq:phaseOP} to write down the quantised Josephson equation.
We begin with a Hamiltonian that describes a finite
superconducting island coupled to a superconducting reservoir,
\begin{eqnarray} \label{eq:H}
\widehat{H} &=&
\widehat{H}_I+\widehat{H}_R+\widehat{H}_C+\widehat{H}_T
\end{eqnarray}
where $\widehat{H}_I$ and $\widehat{H}_R$ are the BCS Hamiltonians
on the island and reservoir respectively, and we introduce
Hamiltonians representing the charging energy of the island,
\begin{eqnarray} \label{eq:hc}
\widehat{H}_C &=& \frac{4e^2}{2C}(\widehat{N}_{I}-n_g)^2
\end{eqnarray}
and the tunnelling between island and reservoir,
\begin{eqnarray} \label{eq:ht+hc}
\widehat{H}_T &=& -T\sum\limits_{k,q}
    {c^{\dag}_{k}c^{\dag}_{-k}c_{-q}c_{q} +
    c^{\dag}_{q}c^{\dag}_{-q}c_{-k}c_{k}}.
\end{eqnarray}
where $T$ is the standard tunnelling matrix element for Cooper Pairs
\cite{wallace}, which we assume for simplicity to be real and
independent of $k,q$. If we make our strong-coupling approximation
and assume that all the electronic energy levels can be considered
equal, we can write these Hamiltonians in terms of spin operators,
as follows:
\begin{eqnarray} \label{eq:hss}
\widehat{H}_{I} &=&  2(\epsilon_{FI}-\mu_I) \widehat{S}_I^Z -
V_I\widehat{S}_I^+
\widehat{S}_1^- \nonumber \\
\widehat{H}_{R} &=&  2(\epsilon_{FR}-\mu_R) \widehat{S}_R^Z -
V_R\widehat{S}_R^+
\widehat{S}_R^- \nonumber \\
\widehat{H}_C &=& \frac{4e^2}{2C}(\widehat{S}_{I}^Z+l_I/2-n_g)^2 \nonumber\\
\widehat{H}_T &=& -T \left(\widehat{S}_I^+ \widehat{S}_R^- +
\widehat{S}_I^- \widehat{S}_R^+ \right).
\end{eqnarray}
Inserting these expressions into equation \ref{eq:H}, we obtain
\begin{eqnarray} \label{eq:hssfinal}
\widehat{H} &=& E_C'(\widehat{S}^Z_I  - n_g')^2 -T
(\widehat{S}^+_I\widehat{S}^-_R +\widehat{S}^-_I\widehat{S}^+_R),
\end{eqnarray}
where we have incorporated the terms from $\widehat{H}_I$ linear and
quadratic in $\widehat{S}^Z_I$ into the renormalised charging energy
and gate voltage represented by $E_C'$ and $n_g'$ respectively. In
the limit that both the reservoir and the island can be considered
infinite, we regain the standard form for the quantised Josephson
junction Hamiltonian,
\begin{eqnarray}
\widehat{H} &=&E_C'\left(i\frac{\partial}{\partial\phi_I} -
n_g'\right)^2
-T\frac{l_{R}l_I}{2} \cos(\phi_I - \phi_R),\nonumber\\
\end{eqnarray}
suggesting that equation \ref{eq:scqu-qmham} can be considered as a
large-size limit where the finite size of the island can be
neglected. However, we are now able to obtain, using equation
\ref{eq:phaseOP}, the next terms in the series expansion,
\begin{eqnarray}\label{eq:finalH}
\widehat{H} &=&E_C'\left(i\frac{\partial}{\partial\phi_I} -
n_g'\right)^2
-T\frac{l_{R}}{2} \bigg{\{} (l_I+1) \cos (\phi_I- \phi_R) \nonumber \\
&&-
\frac{2}{l_I}\left(i\frac{\partial}{\partial\phi_I}-\frac{l_I}{2}\right)^2\cos
(\phi_I- \phi_R)
\nonumber\\
&&- \frac{2}{l_Ii}
\left(i\frac{\partial}{\partial\phi_I}-\frac{l_I}{2}\right)\sin
(\phi_I- \phi_R)+\cdots\bigg\}
\end{eqnarray}
We find that the new terms involve products of both the phase and
the charge operators. Thus the Josephson tunnelling term effectively
depends on the island charge. Making an analogy with a particle in a
potential, with a position corresponding to the phase, we see that
the extra terms in equation \ref{eq:finalH} can be thought of as a
velocity-dependent potential. This effect also breaks the
periodicity of the island energy with $n_g'$, i.e. the energy now
depends on the absolute value of $n_g'$, rather than merely its
value modulo 1. In this derivation we have assumed that the single
electron energies are all equal, and thus the occupations $u_k/v_k$
are all equal. However, these occupations maintain a similar order
of magnitude when different, and so we would expect the extra terms
in eq. \ref{eq:finalH} to be of a similar size when the
strong-coupling approximation is relaxed.

We can make an estimate of the size of these effects for an island
of a given size by calculating $l_I$, the number of electrons within
the cutoff region, by comparing the level spacing to the Debye
energy. For an island with a volume equal to that described in
\cite{nakamura}, we find that $l_I\approx 6\times 10^5$, and thus
the additional terms in equation \ref{eq:finalH} (proportional to
$1/l_I$ are unlikely to be significant. If we consider instead a
nanograin of the type described in
\cite{delftsmallreview,ricrev,RBT}, we find that $l_I \approx 400$,
and thus the extra terms may be relevant.

\section{Conclusions}

We have shown how the quantum Josephson Junction equation, usually
derived by re-quantising the mean field equations of motion, can be
directly derived from a microscopic description of a superconducting
island. We used a simplified Hamiltonian in which the energy of the
individual microscopic electron levels is considered equal that
allowed an exact solution to be found. We have shown how a mean
field approximation leads to a solution that corresponds to a spin
coherent state, which is the BCS state in the appropriate limit. As
well as illustrating how the familiar phenomenology emerges through
the mean field approximation, we showed we can describe effects
beyond the mean field, such as quantum fluctuations. We went on to
rederive the Josephson Junction equation and describe size dependent
corrections to the familiar terms.


\section*{References}

\begin{thebibliography}{99}

\bibitem{josephson} B. D. Josephson, \emph{Phys. Lett.} 1962 {\bf 1} 251

\bibitem{tinkham} M. Tinkham, Introduction to Supeconductivity,(McGraw Hill, New York, 1996)
\bibitem{shnirman97}  A. Shnirman, G. Sch\"on, Z. Hermon  1997, \emph{Phys. Rev. Lett.} {\bf79}
2371
\bibitem{shnirmannature} Yu. Makhlin, G. Sch\"on A. Shnirman, 1999, \emph{Nature}
{\bf398} 305
\bibitem{shnirmanreview} Yu. Makhlin, G. Sch\"on A. Shnirman, 2001, \emph{Rev. Mod. Phys.} {\bf73} 357
\bibitem{younori} J. Q. You and F. Nori, 2005, \emph{Phys. Today} {\bf58} 11

\bibitem{nakamura} Y. Nakamura, Y.A. Pashkin, J.S. Tsai, 1997, \emph{Nature} {\bf 398}, 786
\bibitem{quantronium} D. Vion, A. Assime, A. Cottet, P. Joyez, H. Pothier, C. Urbina, D.Esteve, M.H. Devoret, 2002, \emph{Science}, {\bf 296}  886
\bibitem{nakamura2} Yu. A. Pashkin, T. Yamamoto, O. Astafiev, Y. Nakamura, D. V. Averin, J. S. Tsai
    \emph{Nature}, 2003, {\bf421}, 823
\bibitem{nakamura3} O. Astafiev, Yu. A. Pashkin, T. Yamamoto,  Y. Nakamura, J. S. Tsai
    \emph{Phys. Rev. B.}, 2004, {\bf69}, 180507
\bibitem{delsing} T. Duty, D. Gunnarsson, K. Bladh, P. Delsing
\emph{Phys. Rev. B} 2004 {\bf 69} 140503



\bibitem{jin} X. Y. Jin, J.
Lisenfeld, Y. Koval, A. Lukashenko, A. V. Ustinov, and P. Müller,
2006, \emph{Phys. Rev. Lett.} {\bf96}, 177003

\bibitem{savelev} Sergey Savel'ev, A. L. Rakhmanov, and Franco Nori,
2006, \emph{Phys. Rev. Lett.} {\bf98}, 077002

\bibitem{BCS} J. Bardeen, L.N. Cooper, J.R. Schriefer, 1957, \emph{Phys. Rev.} {\bf108} 1175
\bibitem{richardson} R. W. Richardson, 1963, \emph{Phys. Rev. Lett.} {\bf3}, 277

\bibitem{leeandscully} P. A. Lee and M. O. Scully, 1971, \emph{Phys. Rev. B.} {\bf 3} 769
\bibitem{coherent} R. P. Feynman, R. B. Leighton, M. Sands, Lectures on Physics, Vol. III, {\bf 18}, \emph{Addison-Wesley}, 1965
\bibitem{revival} D. A. Rodrigues, B. L. Gyorffy and T. P. Spiller, 2004 {\emph J. Phys: CM}, {\bf 16} 4477

\bibitem{delftsmallreview} Jan von Delft,  2001, \emph{Ann. Phys. (Leipzig)}, {\bf 10}
1, 2003 \emph{Phys. Rev. B} {\bf 68} 214509
\bibitem{ricrev} J. Dukelsky, S. Pittel and G. Sierra 2004, \emph{Rev. Mod. Phys.} {\bf76}
643

\bibitem{altshuler} E. Yuzbashyan, A. Baytin, B. Altshuler, 2003, \emph{Phys. Rev. B} {\bf68}
214509
\bibitem{wallace} P.R. Wallace and M.J. Stavn, 1965, \emph{Canadian Journal of Physics} {\bf 43} 411

\bibitem{RBT} D.C. Ralph, C.T. Black, M. Tinkham, 1996, \emph{Phys. Rev. Lett.} {\bf76} 688





\end{thebibliography}

\end{document}